\documentclass{article}

\newtheorem{theorem}{Theorem}
\newtheorem{acknowledgement}[theorem]{Acknowledgement}

\input{tcilatex}

\begin{document}

\begin{center}
\bigskip 

\bigskip

\bigskip

\bigskip

{\Large OFF-SHELL BRST-VSUSY \ SUPERALGEBRA }

\bigskip

{\Large FOR D=4 BF THEORIES IN THE }

\bigskip

{\Large SUPERCONNECTION FORMALISM }

\bigskip

\bigskip

\bigskip

{\large A. AIDAOUI}$^{1,2}${\large \ and M. TAHIRI}$^{1}$

$^{1}$\textit{Laboratoire de Physique Th\'{e}orique, Universit\'{e} d'Oran
Es-Senia, 31100 Oran, Algeria}

$^{2}$\textit{Centre Universitaire de B\'{e}char, Institut des Sciences
Exactes, B. P. 417, 08000 B\'{e}char, Algeria}

\textit{hadj\_aidaoui@yahoo.fr}
\end{center}

$\bigskip $

$\bigskip $

$\baselineskip=30pt$

$\centerline{\bf Abstract}\bigskip $

{\normalsize We propose the superconnection formalism to construct the
off-shell BRST-VSUSY superalgebra for D=4 BF theories. The method is based
on the natural introduction of physical fields as well as auxiliary fields
via superconnections and their associated supercurvatures defined on a
superspace. We also give a prescription to build the off-shell BRST-VSUSY
exact quantum action.}

\bigskip


\bigskip


\bigskip

\bigskip

\bigskip

$\bigskip $

{\large It is known that the BRST transformations in Yang-Mills (YM)
theories can be derived systematically from the horizontality conditions
imposed on the supercurvature associated with a superconnection defined on a
(4,2)-dimensional superspace, the so-called BRST superspace, obtained by
extending spacetime with two ordinary anticommuting coordinates }$\left[ 1%
\right] ${\large .\ In such BRST superspace formalism, the gauge fields, the
ghost and antighost fields are interpreted as the lowest components of the
superfield components of the superconnection. This treatment is also
considered in the description for tensor gauge fields }$\left[ 2\right] $%
{\large .}

{\large In Ref.}$\left[ 3\right] ${\large , the authors study the
topological YM theory in the superconnection framework. In such way the
horizontality conditions are modified in order to obtain the BRST
transformation rules of the component fields in a systematic way.}

{\large However, the quantization of BF theories in four dimension has been
discussed within the superfibre bundle formalism }$\left[ 4\right] ${\large %
. In this geometrical framework, the fields present in the quantized theory
have been described through a superconnections introduced over a principal
superfibre bundle. We note that this description is also considered for the
so-called BF-YM theory }$\left[ 5\right] ${\large .}{\Large \ }

{\large In Ref. }$\left[ 6\right] ,${\large \ the author developed the BRST
superspace formalism in order to perform the quantization of BF theories as
a model with reducible gauge symmetry, see e.g. Ref. }$\left[ 7\right] $%
{\large . The method is based on the possibility to enlarging the space of
fields in the quantized theory by auxiliary fields through a supercurvature
of a generalized superconnection. The auxiliary fields are required to
achieve the off-shell nilpotency of the BRST operator. The minimal set of
auxiliary fields is given after having imposed constraints on the
generalized supercurvature in which the consistency with the Bianchi
identities is guaranteed. We note that the same scheme is applied for the
simple supergravity where the classical gauge algebra is open }$\left[ 8%
\right] ${\large .}

{\large It is also known that the introduction of auxiliary fields in the
quantized BF theories guarantees the off-shell closedness of the
superalgebra of the Wess-Zumino type containing besides the BRST symmetry
the vector supersymmetry ( VSUSY ) }$\left[ 9\right] ${\large , which was
already observed in BF theories }$\left[ 10\right] ${\large . We remark that
in the generalized connection formalism as developed in Ref. }$\left[ 9%
\right] ${\large \ (see also Ref. }$\left[ 11\right] ${\large \ ) the VSUSY
is only determined after having built the quantum action.}

{\large Recently,\ the off-shell BRST-VSUSY superalgebra of BF theories in
dimension two }$\left[ 12\right] ${\large \ and four }$\left[ 13\right] $%
{\large \ is considered in terms of superfields defined on the so-called
BRST-VSUSY superspace. In Ref. }$\left[ 13\right] ,${\large \ the latter is
obtained by extending the BRST superspace with four anticommuting
coordinates. In this analysis, the off-shell BRST-VSUSY invariant quantum
action is obtained via a twisting process from supersymmetric theories. }

{\large Furthermore, the off-shell superalgebra of twisted N=2 superYM
theories is constructed by the superconnection formalism of BRST-VSUSY
superspace in two and four dimensions }$\left[ 14\right] ${\large .}

{\large What is still lacking is the construction, in a natural way, of the
BRST-VSUSY superalgebra, for D=4 BF theories, in terms of a superconnection
defined on the BRST-VSUSY superspace. Inspired mainly by the results
obtained in Ref. }$\left[ 6\right] ${\large , we would like to define
superconnections on the BRST-VSUSY superspace, which allow one to derive ab
initio the off-shell BRST-VSUSY superalgebra by using the structure
equations and the Bianchi identities. We mention that, contrary to what
happens in the BRST superspace }$\left[ 6\right] ${\large , the BRST-VSUSY
superspace also leads to the introduction of all auxiliary fields which
implement the BRST-VSUSY exactness of the full quantum action. The latter
can be obtained from a superaction generalizing that constructed in Ref. }$%
\left[ 6\right] ${\large .}

{\large Let us consider the (4,2+4)-dimensional BRST-VSUSY superspace with
local coordinates }$\left( z^{M},\theta ^{\mu }\right) =\left( x^{\mu
},\theta ^{\alpha },\theta ^{\mu }\right) ${\large . Following the guideline
of Ref.}$\left[ 6\right] ${\large \ we introduce over this superspace an
even 1-form superconnection }$\omega ${\large \ and an even 2-form
generalized superconnection }$\phi ${\large . By acting the exterior
covariant superdifferential }$D${\large \ on }$\omega ${\large \ and }$\phi $%
{\large , we obtain the supercurvature }$\Omega ${\large \ ( even 2-form )
and the generalized supercurvature }$\Phi ${\large \ ( even 3-form),
respectively, satisfying the structure equations}

{\large \ }%
\begin{equation}
\Omega =D\omega =d\omega +\frac{1}{2}\left[ \omega ,\omega \right] \text{ },%
\text{ }  \label{eqn1}
\end{equation}%
\[
\]%
\begin{equation}
\Phi =D\phi =d\phi +\left[ \omega ,\phi \right] \text{ },\text{ \ \ \ \ }
\label{eqn2}
\end{equation}%
\[
\]%
{\large and the Bianchi identities}

\begin{equation}
D\Omega =0\text{ },\text{ \ \ \ \ \ \ \ \ \ \ \ \ \ \ \ \ }  \label{eqn3}
\end{equation}%
\[
\]%
\begin{equation}
D\Phi =\left[ \Omega ,\phi \right] \text{ },\text{ \ \ \ \ \ \ \ \ \ }
\label{eqn4}
\end{equation}%
\[
\]%
{\large where }$d${\large \ is the exterior superdifferential and }$\left[ ,%
\right] ${\large \ the graded Lie bracket. The local expressions of the
superconnections and supercurvatures can be written as}%
\[
\]%
{\large \ }%
\begin{equation}
\omega =dz^{M}\omega _{M}+d\theta ^{\mu }\omega _{\mu }^{\prime }\text{ \ },%
\text{ \ \ \ \ \ \ \ \ \ \ \ \ \ \ \ \ \ \ \ \ \ \ \ \ \ \ \ \ \ \ \ \ \ \ \
\ \ \ }  \label{eqn5}
\end{equation}%
\[
\]%
\begin{equation}
\phi =\frac{1}{2}\left\{ dz^{N}dz^{M}\phi _{MN}+dz^{M}d\theta ^{\mu }\phi
_{\mu M}^{\prime }+d\theta ^{\nu }d\theta ^{\mu }\phi _{\mu \nu }^{\prime
\prime }\right\} \text{ },  \label{eqn6}
\end{equation}%
\[
\]%
\begin{equation}
\text{ }\Omega =\frac{1}{2}\left\{ dz^{N}dz^{M}\Omega _{MN}+dz^{M}d\theta
^{\mu }\Omega _{\mu M}^{\prime }+d\theta ^{\nu }d\theta ^{\mu }\Omega _{\mu
\nu }^{\prime \prime }\right\} \text{ },  \label{eqn7}
\end{equation}%
\[
\]%
\begin{eqnarray}
\text{ \ }\Phi &=&\frac{1}{3!}\{dz^{R}dz^{N}dz^{M}\Phi
_{MNR}+dz^{N}dz^{M}d\theta ^{\mu }\Phi _{\mu MN}^{\prime }+  \nonumber \\
&&\text{ \ \ \ \ \ \ }  \nonumber \\
&&dz^{M}d\theta ^{\nu }d\theta ^{\mu }\Phi _{\mu \nu M}^{\prime \prime
}+d\theta ^{\rho }d\theta ^{\nu }d\theta ^{\mu }\Phi _{\mu \nu \rho
}^{\prime \prime \prime }\}\text{ }.\text{ \ }  \label{eq8}
\end{eqnarray}%
\[
\]%
{\large We recall that the components of the superconnections and
supercurvatures are gauge Lie algebra valued superfields. The superfields
components of the supercurvatures are determined by the structure equations.
We have}%
\[
\]%
\begin{equation}
\Omega _{MN}=\partial _{M}\omega _{N}-\left( -1\right) ^{mn}\partial
_{N}\omega _{M}+\left[ \omega _{M},\omega _{N}\right] \text{ },\text{ \ \ \
\ \ \ \ \ \ \ \ \ \ \ \ \ \ \ \ \ \ \ \ \ \ \ \ \ \ \ \ }  \label{eq9}
\end{equation}%
\[
\]%
\begin{equation}
\Omega _{\mu M}^{\prime }=\partial _{\mu }^{\prime }\omega _{M}-\left(
-1\right) ^{m}\partial _{M}\omega _{\mu }^{\prime }+\left[ \omega _{\mu
}^{\prime },\omega _{M}\right] \text{ },\text{ \ \ \ \ \ \ \ \ \ \ \ \ \ \ \
\ \ \ \ \ \ \ \ \ \ \ \ \ \ \ \ \ }  \label{eq10}
\end{equation}%
\[
\]%
\begin{equation}
\Omega _{\mu \nu }^{\prime \prime }=\partial _{\mu }^{\prime }\omega _{\nu
}^{\prime }+\partial _{\nu }^{\prime }\omega _{\mu }^{\prime }+\left[ \omega
_{\mu }^{\prime },\omega _{\nu }^{\prime }\right] \text{ },\text{ \ \ \ \ \
\ \ \ \ \ \ \ \ \ \ \ \ \ \ \ \ \ \ \ \ \ \ \ \ \ \ \ \ \ \ \ \ \ \ \ \ \ \
\ \ \ }  \label{eq11}
\end{equation}%
\[
\]%
\begin{equation}
\Phi _{MNR}=D_{M}\phi _{NR}+\left( -1\right) ^{m\left( n+r\right) }D_{N}\phi
_{RM}+\left( -1\right) ^{r\left( m+n\right) }D_{R}\phi _{MN}\text{ },\text{
\ \ \ }  \label{eq12}
\end{equation}%
\[
\]%
\begin{equation}
\Phi _{\mu MN}^{\prime }=D_{\mu }^{\prime }\phi _{MN}-\left( -1\right)
^{m}D_{M}\phi _{\mu N}^{\prime }+\left( -1\right) ^{n\left( m+1\right)
}D_{N}\phi _{\mu M}^{\prime }\text{ },\text{ \ \ \ \ \ \ \ \ \ \ \ \ }
\label{eq13}
\end{equation}%
\[
\]%
\begin{equation}
\Phi _{\mu \nu M}^{\prime \prime }=D_{\mu }^{\prime }\phi _{\nu M}^{\prime
}+D_{\nu }^{\prime }\phi _{\mu M}^{\prime }+D_{M}\phi _{\mu \nu }^{\prime
\prime }\text{ },\text{ \ \ \ \ \ \ \ \ \ \ \ \ \ \ \ \ \ \ \ \ \ \ \ \ \ \
\ \ \ \ \ \ \ \ \ \ \ \ \ }  \label{eqn14}
\end{equation}%
\[
\]%
\begin{equation}
\Phi _{\mu \nu \rho }^{\prime \prime \prime }=D_{\mu }^{\prime }\phi _{\nu
\rho }^{^{\prime \prime }}+D_{\nu }^{\prime }\phi _{\mu \rho }^{^{\prime
\prime }}+D_{\rho }^{\prime }\phi _{\mu \nu }^{\prime \prime }\text{ },\text{
\ \ \ \ \ \ \ \ \ \ \ \ \ \ \ \ \ \ \ \ \ \ \ \ \ \ \ \ \ \ \ \ \ \ \ \ \ \
\ \ \ \ \ \ }  \label{eqn15}
\end{equation}%
\[
\]%
{\large where }$m=\left\vert z^{M}\right\vert ${\large \ \ is the Grassmann
degree of }$z^{M}${\large , }$D_{M}=\partial _{M}+\left[ \omega _{M},.\right]
${\large , }$D_{\mu }^{\prime }=\partial _{\mu }^{\prime }+\left[ \omega
_{\mu }^{\prime },.\right] ${\large \ and }$\partial _{\mu }^{\prime }=\frac{%
\partial }{\partial \theta ^{\mu }}.${\large \ Similarly, the Bianchi
identities }$\left( 3\right) ${\large \ and }$(4)${\large \ become}%
\[
\]%
\begin{equation}
D_{M}\Omega _{NR}+\left( -1\right) ^{m\left( n+r\right) }D_{N}\Omega
_{RM}+\left( -1\right) ^{r\left( m+n\right) }D_{R}\Omega _{MN}=0\text{ },%
\text{ \ \ \ \ \ \ \ \ \ \ \ }  \label{eqn16}
\end{equation}%
\[
\]%
\begin{equation}
D_{\mu }^{\prime }\Omega _{MN}-\left( -1\right) ^{m}D_{M}\Omega _{\mu
N}^{\prime }+\left( -1\right) ^{n(m+1)}D_{N}\Omega _{\mu M}^{\prime }=0\text{
},\text{ \ \ \ \ \ \ \ \ \ \ \ \ \ \ \ \ \ \ }  \label{eqn17}
\end{equation}%
\[
\]%
\begin{equation}
D_{M}\Omega _{\mu \nu }^{\prime \prime }+D_{\mu }^{\prime }\Omega _{\nu
M}^{\prime }+D_{\nu }^{\prime }\Omega _{\mu M}^{\prime }=0\text{ },\text{ \
\ \ \ \ \ \ \ \ \ \ \ \ \ \ \ \ \ \ \ \ \ \ \ \ \ \ \ \ \ \ \ \ \ \ \ \ \ \
\ \ \ \ \ \ \ }  \label{eqn18}
\end{equation}%
\[
\]%
\begin{equation}
D_{\rho }^{\prime }\Omega _{\mu \nu }^{\prime \prime }+D_{\mu }^{\prime
}\Omega _{\nu \rho }^{\prime \prime }+D_{\nu }^{\prime }\Omega _{\mu \rho
}^{\prime \prime }=0\text{ },\text{ \ \ \ \ \ \ \ \ \ \ \ \ \ \ \ \ \ \ \ \
\ \ \ \ \ \ \ \ \ \ \ \ \ \ \ \ \ \ \ \ \ \ \ \ \ \ \ \ \ }  \label{eqn19}
\end{equation}%
\[
\]%
\[
D_{M}\Phi _{NRS}-\left( -1\right) ^{s\left( m+n+r\right) }D_{S}\Phi
_{MNR}+\left( -1\right) ^{\left( m+n\right) \left( r+s\right) }D_{R}\Phi
_{SMN}-\text{ \ \ \ \ \ \ } 
\]%
\[
\]%
\[
\left( -1\right) ^{m\left( n+r+s\right) }D_{N}\Phi _{RSM}=\left[ \Omega
_{MN},\phi _{RS}\right] +\left( -1\right) ^{s\left( n+r\right) }\left[
\Omega _{MS},\phi _{NR}\right] - 
\]%
\[
\]%
\[
\left( -1\right) ^{nr}\left[ \Omega _{MR},\phi _{NS}\right] -\left(
-1\right) ^{m\left( n+s\right) +sr}\left[ \Omega _{NS},\phi _{MR}\right] + 
\]

\[
\]%
\begin{equation}
\left( -1\right) ^{\left( m+n\right) \left( r+s\right) }\left[ \Omega
_{RS},\phi _{MN}\right] +\left( -1\right) ^{m\left( n+r\right) }\left[
\Omega _{NR},\phi _{MS}\right] \text{ },  \label{eqn20}
\end{equation}%
\[
\]%
\[
D_{M}\Phi _{\mu NR}^{\prime }-\left( -1\right) ^{m(n+r+1)}D_{\mu }^{\prime
}\Phi _{NRM}+\left( -1\right) ^{r\left( m+n+1\right) +m}D_{R}\Phi _{\mu
MN}^{\prime }+\text{ \ \ \ \ \ \ \ \ \ \ \ \ }
\]%
\[
\]%
\[
\left( -1\right) ^{m(n+r+1)+n}D_{N}\Phi _{\mu RM}^{\prime }=-\left(
-1\right) ^{n}\left[ \Omega _{MN},\phi _{\mu R}^{\prime }\right] -\left(
-1\right) ^{r(m+n)+m}\left[ \Omega _{\mu R}^{\prime },\phi _{MN}\right] 
\]%
\[
\]%
\[
-\left( -1\right) ^{m}\left[ \Omega _{\mu M}^{\prime },\phi _{NR}\right]
+\left( -1\right) ^{m(n+1)}\left[ \Omega _{\mu N}^{\prime },\phi _{MR}\right]
-
\]%
\[
\]%
\begin{equation}
\left( -1\right) ^{(r+n)(m+1)+m}\left[ \Omega _{NR},\phi _{\mu M}^{\prime }%
\right] +\left( -1\right) ^{r(n+1)}\left[ \Omega _{MR},\phi _{\mu N}^{\prime
}\right] \text{ },  \label{eqn21}
\end{equation}%
\[
\]%
\[
D_{M}\Phi _{\nu \mu N}^{\prime \prime }+D_{\mu }^{\prime }\Phi _{\nu
MN}^{\prime }+D_{\nu }^{\prime }\Phi _{\mu MN}^{\prime }-\left( -1\right)
^{mn}D_{N}\Phi _{\nu \mu M}^{\prime \prime }=\text{ \ \ \ \ \ \ \ \ \ \ \ \
\ \ \ \ \ \ \ \ \ \ \ \ \ \ \ \ \ \ \ \ \ \ \ \ \ \ \ }
\]%
\[
\]%
\[
\left[ \Omega _{MN},\phi _{\nu \mu }^{\prime \prime }\right] -\left(
-1\right) ^{m}\left[ \Omega _{\nu M}^{\prime },\phi _{\mu N}^{\prime }\right]
+\left[ \Omega _{\nu \mu }^{\prime \prime },\phi _{MN}\right] +\left(
-1\right) ^{n(m+1)}\left[ \Omega _{\mu N}^{\prime },\phi _{\nu M}^{\prime }%
\right] -
\]%
\[
\]%
\begin{equation}
\left( -1\right) ^{m}\left[ \Omega _{\mu M}^{\prime },\phi _{\nu N}^{\prime }%
\right] +\left( -1\right) ^{n(m+1)}\left[ \Omega _{\nu N}^{\prime },\phi
_{\mu M}^{\prime }\right] \text{ },  \label{eqn22}
\end{equation}%
\[
\]%
\[
\left( -1\right) ^{m}D_{M}\Phi _{\rho \nu \mu }^{\prime \prime \prime
}-D_{\mu }^{\prime }\Phi _{\rho \nu M}^{\prime \prime }-D_{\nu }^{\prime
}\Phi _{\mu \rho M}^{\prime \prime }-D_{\rho }^{\prime }\Phi _{\nu \mu
M}^{\prime \prime }=-\left[ \Omega _{\nu \mu }^{\prime \prime },\phi _{\rho
M}^{\prime }\right] -\left[ \Omega _{\rho \nu }^{\prime \prime },\phi _{\mu
M}^{\prime }\right] 
\]%
\[
\]%
\begin{equation}
-\left[ \Omega _{\rho M}^{\prime },\phi _{\nu \mu }^{\prime \prime }\right] -%
\left[ \Omega _{\mu M}^{\prime },\phi _{\rho \nu }^{\prime \prime }\right] -%
\left[ \Omega _{\nu M}^{\prime },\phi _{\rho \mu }^{\prime \prime }\right] -%
\left[ \Omega _{\rho \mu }^{\prime \prime },\phi _{\nu M}^{\prime }\right] ,
\label{eqn23}
\end{equation}%
\[
\]%
\[
D_{\sigma }^{\prime }\Phi _{\rho \nu \mu }^{\prime \prime \prime }+D_{\mu
}^{\prime }\Phi _{\sigma \rho \nu }^{\prime \prime \prime }+D_{\nu }^{\prime
}\Phi _{\mu \sigma \rho }^{\prime \prime \prime }+D_{\rho }^{\prime }\Phi
_{\nu \mu \sigma }^{\prime \prime \prime }=\left[ \Omega _{\sigma \nu
}^{\prime \prime },\phi _{\rho \mu }^{\prime \prime }\right] +\left[ \Omega
_{\rho \mu }^{\prime \prime },\phi _{\sigma \nu }^{\prime \prime }\right] 
\text{ \ \ \ \ \ \ \ \ \ \ \ \ \ }
\]%
\[
\]%
\begin{equation}
\left[ \Omega _{\sigma \rho }^{\prime \prime },\phi _{\nu \mu }^{\prime
\prime }\right] +\left[ \Omega _{\mu \sigma }^{\prime \prime },\phi _{\rho
\nu }^{\prime \prime }\right] +\left[ \Omega _{\nu \mu }^{\prime \prime
},\phi _{\sigma \rho }^{\prime \prime }\right] +\left[ \Omega _{\rho \nu
}^{\prime \prime },\phi _{\mu \sigma }^{\prime \prime }\right] \text{ }.
\label{eqn24}
\end{equation}%
\[
\]%
{\large To translate the above equations into equations determining the
off-shell BRST-VSUSY superalgebra of D=4 BF theories, we should interpret
the fields occuring in such theories geometrically. We note that the
physical fields and the auxiliary fields are introduced through the
superconnections and the supercurvatures, respectively. The auxiliary fields
are required for the construction of the off-shell structure of the theory.
To this end, we will be only interested to the superfields }$\omega _{M}$%
{\large \ and }$\phi _{MN}${\large \ \ which give the physical fields as
constructed in Ref.}$\left[ 6\right] ${\large , i.e. we can put}

\begin{equation}
\omega _{\mu }^{\prime }=0\text{ },\ \ \ \ \ \ \ \phi _{\mu M}^{\prime }=0%
\text{ },\ \ \ \ \ \ \ \phi _{\mu \nu }^{\prime \prime }=0.  \label{eq25}
\end{equation}%
\[
\]%
{\large At this point, we remark that the lowest components }$\omega _{\mu
}\mid ,${\large \ }$\omega _{1}\mid ,${\large \ }$\omega _{2}\mid ${\large \
, }$\partial _{1}\omega _{2}\mid ${\large \ and }$\Omega _{\mu \nu }\mid $%
{\large \ describe as usual the gauge potential }$A_{\mu }^{0}${\large , the
ghost }$A_{0}^{1}${\large , the anti-ghost }$A_{0}^{-1}${\large \ , the
Stueckelberg field }$b_{0}^{0}${\large \ and the YM curvature }$F_{\mu \nu
}^{0}${\large , respectively. The symbol }$"\mid "${\large \ \ indicates
that the quantity is evaluated at }$\theta ^{\alpha }=0${\large \ and }$%
\theta ^{\mu }=0${\large \ . Here and in what follows, we have used the
convention that a field }$X$ \ {\large with Lorentz degree }$p$ {\large and
ghost number }$q${\large \ is denoted }$X_{\mu _{1}...\mu _{p}}^{q}${\large %
\ ( or }$X_{p}^{q}$){\large . Besides the }$A_{\mu }^{0}${\large \ system of
fields related to the YM symmetry, i.e. }$\left\{ A_{\mu }^{0}{\large ,\ }%
A_{0}^{1}{\large ,\ }A_{0}^{-1},{\large \ }b_{0}^{0}\right\} ${\large ,
there exists the system of fields related to the reducible symmetry and
associated with the rank-two tensor gauge field }$B_{\mu \nu }^{0}${\large .
This is identified as follows: }$B_{\mu \nu }^{0}=\phi _{\mu \nu }\mid $%
{\large , }$B_{\mu }^{1}=\phi _{\mu 1}\mid ${\large \ is the ghost
associated with }$B_{\mu \nu }^{0}${\large , }$B_{\mu }^{-1}=\phi _{\mu
2}\mid ${\large \ is the anti-ghost of }$B_{\mu }^{1}${\large \ , }$\pi
_{\mu }^{0}=\partial _{1}\phi _{\mu 2}\mid ${\large \ is the associated
Stueckelberg field, }$B_{0}^{2}=\phi _{11}/2\mid $ {\large is the ghost for
the ghost }$B_{\mu }^{1}${\large , }$B_{0}^{-2}=\phi _{22}/2\mid ${\large \
is the anti-ghost of }$B_{0}^{2}$ {\large , }$\pi _{0}^{-1}=\partial
_{1}\phi _{22}\mid ${\large \ is the associated Stueckelberg field, }$K_{\mu
\nu \rho }^{0}=\Phi _{\mu \nu \rho }\mid ${\large \ is the curvature of }$%
B_{\mu \nu }^{0}${\large \ and }$\left( B_{0}^{0}=\phi _{12}\mid {\large ,\ }%
\pi _{0}^{1}=\partial _{1}\phi _{12}\mid \right) ${\large \ is a pair of
fields which takes into account a further degeneracy associated with }$\pi
_{\mu }^{0}${\large . }

{\large The next step is to search for constraints to the supercurvatures in
which the consistency with the Bianchi identities is ensured. This
requirement guarantees then the off-shell closedness of the BRST-VSUSY
superalgebra thanks to the structure equations and the Bianchi identities.
First, we note that the BRST transformations are obtained by imposing as
usual the following constraints}

\begin{equation}
\Omega _{\mu \alpha }=\Omega _{\alpha \beta }=0\text{ },\ \text{\ \ \ \ }%
\Phi _{\mu \nu 2}=\Phi _{\mu 22}=\Phi _{\mu 12}=\Phi _{222}=0.  \label{eqn26}
\end{equation}%
\[
\]%
{\large It is easy to see through an analysis of the Bianchi identities }$%
\left( 16\right) ${\large \ and }$\left( 20\right) ${\large \ that the
constraints }$\left( 26\right) ${\large \ fulfill the consistency with these
equations. However, it is worth noting that the remaining supercurvature
components }$\Phi _{\mu \nu 1}${\large , }$\Phi _{1\mu 1}${\large , }$\Phi
_{111}${\large \ permit us to introduce auxiliary fields defined by}

\begin{equation}
\Phi _{\mu \nu 1}\mid =E_{\mu \nu }^{1}\text{ },\text{ \ \ \ \ \ \ \ }\frac{1%
}{2}\text{\ }\Phi _{1\mu 1}\mid =E_{\mu }^{2}\text{ \ , \ \ \ \ \ \ \ \ \ \ }%
\frac{1}{6}\text{\ }\Phi _{111}\mid =E_{0}^{3}\text{ , \ \ \ \ \ \ \ \ \ \ \
\ \ }  \label{eqn27}
\end{equation}%
\[
\]%
{\large which are required for the construction of the off-shell nilpotent
BRST transformations (see Ref.}$\left[ 6\right] ${\large ). The superfields }%
$\Phi _{\mu \nu 1}${\large , }$\Phi _{1\mu 1}${\large , }$\Phi _{111}$%
{\large \ can be expanded in power series of }$\theta ^{\mu }${\large \ \ as
follows}

{\large \ }%
\[
\Phi _{\mu \nu 1}=E_{\mu \nu }^{1}\left( z\right) +\theta ^{\rho }E_{\mu \nu
\rho }^{0}\left( z\right) +\frac{1}{2!}\theta ^{\sigma }\theta ^{\rho
}E_{\mu \nu \rho \sigma }^{-1}\left( z\right) ,\text{ \ \ \ \ \ \ \ \ \ \ \
\ \ \ \ \ \ \ \ \ \ \ \ \ \ \ \ \ \ \ \ \ \ \ \ \ \ } 
\]%
\ 
\[
\text{\ }\frac{1}{2}\Phi _{1\mu 1}=E_{\mu }^{2}\left( z\right) +\theta ^{\nu
}E_{\mu \nu }^{1}\left( z\right) +\frac{1}{2!}\theta ^{\rho }\theta ^{\nu
}E_{\mu \nu \rho }^{0}\left( z\right) +\text{ }\frac{1}{3!}\theta ^{\sigma
}\theta ^{\rho }\theta ^{\nu }E_{\mu \nu \rho \sigma }^{-1}\left( z\right) ,%
\text{\ \ \ \ \ \ \ \ \ \ \ \ \ \ \ \ \ \ \ } 
\]

\[
\frac{1}{6}\Phi _{111}=E_{0}^{3}\left( z\right) +\theta ^{\mu }E_{\mu
}^{2}\left( z\right) +\frac{1}{2!}\theta ^{\nu }\theta ^{\mu }E_{\mu \nu
}^{1}\left( z\right) +\frac{1}{3!}\theta ^{\rho }\theta ^{\nu }\theta ^{\mu
}E_{\mu \nu \rho }^{0}\left( z\right) +\text{\ \ \ \ \ \ \ \ \ \ \ \ \ \ \ \
\ \ \ \ \ \ \ \ \ \ } 
\]%
\ 
\begin{equation}
\frac{1}{4!}\theta ^{\sigma }\theta ^{\rho }\theta ^{\nu }\theta ^{\mu
}E_{\mu \nu \rho \sigma }^{-1}\left( z\right) ,  \label{eqn28}
\end{equation}%
\[
\]%
{\large where the set of auxiliary fields }$\left\{ E_{\mu \nu }^{1},E_{\mu
}^{2},E_{0}^{3},E_{\mu \nu \rho }^{0},E_{\mu \nu \rho \sigma }^{-1}\right\} $%
{\large , which is obtained by evaluating the superfields components in the
development }$\left( 28\right) ${\large \ at }$\theta ^{\alpha }=0$, {\large %
are needed for the construction of the off-shell structure of D=4 BF
theories.}

{\large Obviously, from }$\left( 10\right) ${\large , }$\left( 11\right) $%
{\large , }$\left( 13\right) ${\large , }$\left( 14\right) ${\large \ and }$%
\left( 15\right) ${\large \ and in view of }$\left( 25\right) ${\large , it
follows that}

\begin{equation}
\Omega _{\mu \nu }^{\prime \prime }=0\text{ },\text{ \ \ \ \ \ \ }\Phi _{\mu
\nu M}^{\prime \prime }=\Phi _{\mu \upsilon \rho }^{\prime \prime \prime }=0%
\text{ },\text{ \ \ \ \ }  \label{eqn29}
\end{equation}%
\ 

\begin{equation}
\Omega _{\mu M}^{\prime }=\partial _{\mu }^{\prime }\omega _{M}\text{ },%
\text{ \ \ \ \ \ \ \ \ \ \ \ \ \ \ \ \ \ \ \ \ \ \ \ \ \ \ \ \ }
\label{eqn30}
\end{equation}

\begin{equation}
\Phi _{\mu MN}^{\prime }=\partial _{\mu }^{\prime }\phi _{MN}\text{ },\text{
\ \ \ \ \ \ \ \ \ \ \ \ \ \ \ \ \ \ \ \ \ \ \ \ \ }  \label{eqn31}
\end{equation}%
\[
\]%
\ {\large which satisfy automatically the consistency of the Bianchi
identities }$\left( 18\right) ${\large , }$\left( 19\right) ${\large , }$%
\left( 22\right) ${\large , }$\left( 23\right) ${\large \ and }$\left(
24\right) ${\large .}

{\large Moreover, through an analysis of the Bianchi identity }$\left(
17\right) ${\large , we deduce the following solutions}%
\begin{equation}
\ \Omega _{\mu 1}^{\prime }=\omega _{\mu }\text{ \ },\ \ \ \ \ \ \ \ \ \ \
\Omega _{\mu 2}^{\prime }=0.  \label{eqn32}
\end{equation}%
{\large These permit us to introduce an auxiliary field given by}

\begin{equation}
\Omega _{\mu \nu }^{\prime }\mid =H_{\mu \nu }^{-1}\text{ },\text{ \ \ }
\label{eq33}
\end{equation}%
\[
\]%
{\large which is required for the construction of off-shell BRST-VSUSY
transformations. The superfield }$\Omega _{\mu \nu }^{\prime }${\large \ can
be expanded in power series of }$\theta ^{\mu }${\large \ as follows}

\begin{equation}
\Omega _{\mu \nu }^{\prime }=H_{\mu \nu }^{-1}\left( z\right) +\theta ^{\rho
}H_{\mu \nu \rho }^{-2}\left( z\right) +\frac{1}{2!}\theta ^{\rho }\theta
^{\sigma }H_{\mu \nu \sigma \rho }^{-3}\left( z\right) {\large .}
\label{eq34}
\end{equation}%
\[
\]%
{\large The set of auxiliary} {\large \ fields }$\left\{ H_{\mu \nu
}^{-1},H_{\rho \mu \nu }^{-2},H_{\sigma \rho \mu \nu }^{-3}\right\} ${\large %
, which is obtained by evaluating the superfields components in the
development }$\left( 34\right) ${\large \ at }$\theta ^{\alpha }=0${\large ,
is also needed for the construction of the off-shell structure of D=4 BF
theories.}

{\large Turning now to the situation of the superfields }$\Phi _{\mu
MN}^{\prime }$\ {\large , a similar analysis of the Bianchi identity }$%
\left( 21\right) ${\large \ leads to the following solutions}

{\large \ }

\begin{equation}
\Phi _{\mu \nu 1}^{\prime }=\phi _{\mu \nu }\text{ , \ \ \ }\frac{1}{2}\Phi
_{\mu 11}^{\prime }=\phi _{\mu 1}\text{ },\text{\ \ \ \ }\Phi _{\mu
22}^{\prime }=\Phi _{\mu 12}^{\prime }=\Phi _{\mu 21}^{\prime }=0.\text{\ \
\ \ }  \label{eqn35}
\end{equation}%
\[
\]%
{\large Again, the remaining superfields }${\large \ }\Phi _{\mu \nu \rho
}^{\prime }${\large \ and }${\large \ }\Phi _{\mu \nu 2}^{\prime }${\large \
as given in }$\left( 31\right) ${\large \ ensure automatically the
consistency of the Bianchi identities and in order to find exactly the
off-shell BRST-VSUSY transformations, we can choose}

\begin{equation}
\Phi _{\mu \nu \rho }^{\prime }\mid =\varepsilon _{\mu \nu \rho \sigma
}\partial ^{\sigma }A_{0}^{-1},\text{ \ \ \ \ \ \ \ \ \ \ \ }\Phi _{\mu \nu
2}^{\prime }\mid =-\eta _{\mu \nu }B_{0}^{-2},  \label{eqn36}
\end{equation}%
\[
\]%
{\large \ where }$\varepsilon _{\mu \nu \rho \sigma }${\large \ is the
antisymmetric Levi-civita tensor and }$\eta _{\mu \nu }$\ {\large is the
flat Minkowskian metric in 4-dimensions. }

{\large \ However, in order to obtain the off-shell BRST-VSUSY
transformations of all the fields, we note that the BRST operator is related
as usual to the partial derivative }$\partial _{1}$ $\left[ 6\right] $%
{\large , whereas the VSUSY\ operator can be related to the differential
operator }$D_{\theta ^{\mu }}=\partial _{\mu }^{\prime }+\theta \partial
_{\mu }${\large . So, we can make the following substitution }

\[
\Omega _{\mu M}^{\prime }\longrightarrow \Lambda _{\mu M}=\Omega _{\mu
M}^{\prime }+\theta \partial _{\mu }\omega _{M}\text{ }, 
\]

\begin{equation}
\Phi _{\mu MN}^{\prime }\longrightarrow \Pi _{\mu MN}=\Phi _{\mu MN}^{\prime
}+\theta \partial _{\mu }\phi _{MN}\text{ },  \label{eqn37}
\end{equation}

\[
\]%
{\large where, we have remplaced }$\partial _{\mu }^{\prime }${\large \ by }$%
D_{\theta ^{\mu }}${\large \ in the superfield components\ }$\Omega _{\mu
M}^{\prime }${\large \ and }$\Phi _{\mu MN}^{\prime }${\large .}

{\large Then, we realize the following identifications}

\begin{equation}
Q\left( \psi \mid \right) =\left( \partial _{1}\psi \right) \mid {\large ,}
\label{eq38}
\end{equation}

\begin{equation}
Q_{\mu }\left( \psi \mid \right) =\left( D_{\theta ^{\mu }}\psi \right) \mid
,  \label{eq39}
\end{equation}%
\[
\]%
{\large where }$Q${\large \ and }$Q_{\mu }${\large \ are interpreted as the
BRST and VSUSY operator for D=4 BF theories, respectively, and }$\psi $%
{\large \ is any superfield with the replacements }$\left( 37\right) $%
{\large .}

{\large Furthermore, from the Bianchi identity }$\left( 21\right) ${\large \
and in view of }$\left( 37\right) ,${\large \ we deduce that}

\[
E_{\mu \nu \rho }^{0}=-\varepsilon _{\mu \nu \rho \sigma }\partial ^{\sigma
}b_{0}^{0}-\varepsilon _{\mu \nu \rho \sigma }\left[ A_{0}^{1},\partial
^{\sigma }A_{0}^{-1}\right] -K_{\mu \nu \rho }^{0}+\left[ H_{\mu \nu
}^{-1},B_{\rho }^{1}\right] -\left[ H_{\mu \rho }^{-1},B_{\nu }^{1}\right] , 
\]

\ \ 
\begin{eqnarray}
E_{\lambda \mu \nu \rho }^{-1} &=&-\left[ H_{\lambda \mu }^{-1},B_{\nu \rho
}^{0}\right] -\left[ H_{\lambda \nu }^{-1},B_{\rho \mu }^{0}\right] -\left[
H_{\lambda \rho }^{-1},B_{\mu \nu }^{0}\right] -\left[ H_{\mu \nu
}^{-1},B_{\lambda \rho }^{0}\right] +\text{ \ \ \ \ \ \ \ \ \ \ }  \nonumber
\\
\text{ \ \ \ \ \ \ \ \ } &&\left[ H_{\mu \rho }^{-1},B_{\lambda \nu }^{0}%
\right] +\left[ H_{\mu \nu \lambda }^{-2},B_{\rho }^{1}\right] -\left[
H_{\mu \rho \lambda }^{-2},B_{\nu }^{1}\right] .  \label{eqn40}
\end{eqnarray}%
\[
\]%
{\large In the end, by inserting }$\left( 26\right) ,\left( 28\right)
,\left( 29\right) ,\left( 32\right) ,\left( 34\right) ,\left( 35\right) $%
{\large \ and }$\left( 36\right) ${\large \ in the structure equations and
the Bianchi identities, and by using the identification }$\left( 38\right) $%
{\large , we deduce the off-shell BRST\ transformations of the auxiliary
fields and the original fields as given in Ref. }$\left[ 6\right] ${\large \
( see also Ref. }$\left[ 9\right] ${\large \ )}

{\large \ \ } 
\[
QA_{0}^{1}=-\frac{1}{2}\left[ A_{0}^{1},A_{0}^{1}\right] \text{ \ , \ \ \ \
\ \ \ \ \ \ \ }QA_{\mu }^{0}=D_{\mu }A_{0}^{1}\text{ ,} 
\]%
\ \ \ \ \ \ \ \ \ \ \ \ \ \ \ 

\[
\ QB_{\mu \nu }^{0}=-(D_{\mu }B_{\nu }^{1}-D_{\nu }B_{\mu }^{1})-\left[
A_{0}^{1},B_{\mu \nu }^{0}\right] +E_{\mu \nu }^{1}, 
\]%
\ \ \ \ \ \ \ \ \ \ \ \ \ \ \ \ \ 

\[
QB_{\mu }^{1}=D_{\mu }B_{0}^{2}-\left[ A_{0}^{1},B_{\mu }^{1}\right] +E_{\mu
}^{2}\text{ \ , \ }QB_{0}^{2}=-\left[ A_{0}^{1},B_{0}^{2}\right] +E_{0}^{3}%
\text{ \ ,} 
\]

\[
QH_{\mu \nu }^{-1}=F_{\mu \nu }-\left[ A_{0}^{1},H_{\mu \nu }^{-1}\right] 
\text{ \ ,} 
\]

\[
QH_{\mu \nu \rho }^{-2}=-\tsum_{\left( \mu \nu \rho \right) }D_{\mu }H_{\nu
\rho }^{-1}-\left[ A_{0}^{1},H_{\mu \nu \rho }^{-2}\right] \text{ \ ,} 
\]%
\ 

\[
QH_{\mu \nu \rho \sigma }^{-3}=\tsum_{\left( \mu \nu \rho \sigma \right)
}D_{\mu }H_{\nu \rho \sigma }^{-2}-\left[ A_{0}^{1},H_{\mu \nu \rho \sigma
}^{-3}\right] +\frac{1}{2}\sum_{\left( \mu \nu \rho \sigma \right) }\left[
H_{\mu \nu }^{-1},H_{\rho \sigma }^{-1}\right] \ , 
\]%
\ \ \ \ \ \ \ \ \ \ \ \ \ \ \ \ \ \ \ \ \ \ 

\[
QE_{\mu \nu }^{1}=\sum_{\left( \mu \nu \right) }D_{\mu }E_{\nu }^{2}-\left[
A_{0}^{1},E_{\mu \nu }^{1}\right] +\left[ F_{\mu \nu }^{0},B_{0}^{2}\text{\ }%
\right] \text{\ ,} 
\]

\[
QE_{\mu }^{2}=D_{\mu }E_{0}^{2}-\left[ A_{0}^{1},E_{\mu }^{2}\right] \text{
, \ \ \ \ \ }QE_{0}^{3}=-\left[ A_{0}^{1},E_{0}^{3}\right] \text{ ,} 
\]

\[
QA_{0}^{-1}=b_{0}^{0}\text{ \ , \ \ \ \ \ \ \ \ \ \ \ \ \ }Qb_{0}^{0}=0\text{
,} 
\]

\[
QB_{\mu }^{-1}=\pi _{\mu }^{0}\text{ \ , \ \ \ \ \ \ \ \ \ \ }Q\pi _{\mu
}^{0}=0\text{ \ ,} 
\]

\[
QB_{0}^{-2}=\pi _{0}^{-1}\text{ \ , \ \ \ \ \ \ \ \ \ }Q\pi _{0}^{-1}=0\text{
\ ,} 
\]

\begin{equation}
QB_{0}^{0}=\pi _{0}^{1}\text{ \ , \ \ \ \ \ \ \ \ \ \ }Q\pi _{0}^{1}=0\text{
\ ,}  \label{eqn41}
\end{equation}%
\[
\]%
{\large where }$\dsum\limits_{\left( MN..\right) }$ {\large means a cyclic
sum over M, N,... .}

{\large Then, by using }$\left( 39\right) ${\large , we obtaine the
following VSUSY transformations as given in Ref.}$\left[ 9\right] ${\large .}

\[
\]%
\[
Q_{\mu }A_{0}^{1}=A_{\mu }^{0}\text{ , \ \ \ \ \ \ }Q_{\mu }A_{\nu
}^{0}=H_{\mu \nu }^{-1}\text{ , \ \ \ \ \ }Q_{\mu }H_{\nu \rho }^{-1}=H_{\nu
\rho \mu }^{-2}\text{ , } 
\]

\[
\text{\ }Q_{\mu }H_{\nu \rho \sigma }^{-2}=H_{\nu \rho \mu \sigma }^{-3}%
\text{ , \ \ \ \ \ }Q_{\mu }H_{\nu \rho \sigma \tau }^{-3}=0\text{ , \ } 
\]

\[
Q_{\mu }B_{0}^{2}=B_{\mu }^{1}\ ,\ \ \ \ \ \ \ Q_{\mu }B_{\nu }^{1}=B_{\mu
\nu }^{0},\ \ \ \ \ \ Q_{\mu }B_{\nu \rho }^{0}=\varepsilon _{\mu \nu \rho
\sigma }^{{}}\partial ^{\sigma }A_{0}^{-1}, 
\]

\[
Q_{\mu }E_{0}^{3}=E_{\mu }^{2}\text{ \ },\text{ \ \ \ \ \ \ \ \ \ \ \ }%
Q_{\mu }E_{\nu }^{2}=E_{\nu \mu }^{1}\text{ \ },\text{ \ \ \ \ \ \ \ \ }%
Q_{\mu }E_{\nu \rho }^{1}=E_{\nu \rho \mu }^{0}\ , 
\]

\[
\text{\ }Q_{\mu }E_{\nu \rho \sigma }^{0}=E_{\nu \rho \mu \sigma }^{-1}\ ,\
\ \ \ \ \ \ \ Q_{\mu }E_{\nu \rho \sigma \lambda }^{-1}=0, 
\]

\[
Q_{\mu }A_{0}^{-1}=0\text{ \ , \ \ \ \ \ \ \ \ \ }Q_{\mu }b_{0}^{0}=\partial
_{\mu }A_{0}^{-1}\text{ \ ,} 
\]

\[
Q_{\mu }B_{\nu }^{-1}=-\eta _{\mu \nu }B_{0}^{-2}\text{ \ , \ \ \ \ \ \ }%
Q_{\mu }\pi _{\nu }^{0}=\partial _{\mu }B_{\nu }^{-1}+\eta _{\mu \nu }\pi
_{0}^{-1}\text{ \ , } 
\]

\[
Q_{\mu }B_{0}^{-2}=0\text{ \ , \ \ \ \ \ \ \ \ \ \ \ }Q_{\mu }\pi
_{0}^{-1}=\partial _{\mu }B_{0}^{-2}\text{ \ , } 
\]

\begin{equation}
Q_{\mu }B_{0}^{0}=0\text{ \ , \ \ \ \ \ \ \ \ \ \ \ }Q_{\mu }\pi
_{0}^{1}=\partial _{\mu }B_{0}^{0}\text{ \ .}  \label{eqn42}
\end{equation}

\[
\]%
{\large It is easy to see that the BRST-VSUSY superalgebra }

\begin{equation}
\left\{ Q,Q\right\} =0\text{ \ , \ \ \ }\left\{ Q_{\mu },Q_{\nu }\right\} =0%
\text{ \ , \ \ \ \ }\left\{ Q,Q_{\mu }\right\} =\partial _{\mu }\text{ \ \ \
\ }  \label{eq43}
\end{equation}%
\[
\]%
{\large is automatically satisfied off-shell, thanks to the structure
equations and the Bianchi identities.}

{\large Finally, in order to arrive at the off-shell BRST-VSUSY quantum
action of D=4 BF theories, let us recall that the gauge fixing action of D=4
BF theories in the BRST superspace formalism, is constructed from a gauge
fixing superaction given by }$\left[ 6\right] $

\[
S_{sgf}=\partial _{1}\omega _{2}\partial ^{\mu }\omega _{\mu }-\omega
_{2}\partial ^{\mu }\partial _{1}\omega _{\mu }+\partial _{1}\phi _{2}^{\mu
}\partial ^{\nu }\phi _{\nu \mu }-\phi _{2}^{\mu }\partial ^{\nu }\partial
_{1}\phi _{\nu \mu }+\partial _{1}\phi _{1}^{\mu }\partial _{\mu }\phi _{22}-%
\text{ \ \ \ \ \ \ \ } 
\]%
\[
\phi _{1}^{\mu }\partial _{\mu }\partial _{1}\phi _{22}+\partial _{1}\phi
_{2}^{\mu }\partial _{\mu }\phi _{12}-\phi _{2}^{\mu }\partial _{\mu
}\partial _{1}\phi _{12}+\partial _{1}\phi _{22}\partial _{1}\phi _{12} 
\]%
\[
\]%
\begin{equation}
=\partial _{1}\left[ \omega _{2}\partial ^{\mu }\omega _{\mu }+\phi
_{2}^{\mu }\partial ^{\nu }\phi _{\nu \mu }+\phi _{1}^{\mu }\partial _{\mu
}\phi _{22}+\phi _{2}^{\mu }\partial _{\mu }\phi _{12}+\phi _{22}\partial
_{1}\phi _{12}\right] .  \label{eqn44}
\end{equation}%
\[
\]%
{\large In the BRST-VSUSY superspace formalism, }$\left( 44\right) ${\large %
\ can be written, up to a total divergence, as}

\begin{equation}
S_{sgf}=-\partial _{1}D_{\theta ^{\mu }}(\omega _{2}\partial ^{\mu }\omega
_{1}+\phi _{2}^{\mu }\partial ^{\nu }\phi _{\nu 1}+\phi _{2}^{\mu }\partial
_{1}\phi _{12}).  \label{eqn45}
\end{equation}%
\[
\]%
{\large Furthermore, in order to find the classical action, we define the
extented classical superaction as follows}

\begin{equation}
S_{s}^{0}=\frac{1}{4}\varepsilon ^{\mu \nu \rho \sigma }\partial
_{1}D_{\theta ^{\sigma }}(\Omega _{\mu \nu }^{\prime }\phi _{\rho 1}+\frac{1%
}{6}\partial _{\rho }^{\prime }\Omega _{\mu \nu }^{\prime }\phi _{11}).
\label{eq46}
\end{equation}%
\[
\]%
{\large \ It is easy to see that the full quantum action }$S_{q}${\large \
defined by}%
\begin{equation}
S_{q}=\left( S_{s}^{0}+S_{sgf}\right) \mid   \label{eqn47}
\end{equation}%
{\large is BRST-VSUSY exact. It can be also written in the BRST exact form
leading to the same quantum action with the Landau gauge obtained in the
context of the generalized connection formalism }$\left[ 9\right] ${\large .}

{\large In conclusion, working with the same spirit as in Ref.}$[6]${\large %
, we have performed the quantization of D=4 BF theories by using the
superconnection formalism. In this analysis, the }$A_{\mu }^{0}${\large \
and }$B_{\mu \nu }^{0}${\large \ systems of fields are described as usual
through even superconnections over a BRST-VSUSY superspace. Using the
exterior covariant superdifferential on the superconnections gives even
supercurvatures, which lead to the determination of the off-shell BRST-VSUSY
transformations. The off-shell closedness is realized by introducing through
the supercurvatures two minimal sets of auxiliary fields required for the
consistency of the BRST-VSUSY superspace geometry. Finally, in order to find
the full quantum action, we have built the superaction generalizing that in
Ref.}$\left[ 6\right] ${\large . This results in the construction of the
off-shell BRST-VSUSY exact quantum action of D=4 BF theories.}

\bigskip

\begin{acknowledgement}
\bigskip MT would like to thank the Alexander von Humboldt Stiftung for
support
\end{acknowledgement}

\begin{quote}
\bigskip

\bigskip

\bigskip

\bigskip

\bigskip 
\end{quote}

\end{document}